# Vacuum friction on a rotating pair of atoms

Hervé Bercegol [•,*] & Roland Lehoucq [▫]

[•] Service de Physique de l'Etat Condensé, DSM/IRAMIS/SPEC/SPHYNX, CNRS UMR 3680, CEA Saclay, F-91191 Gif-sur-Yvette, France

[▫] Laboratoire AIM-Paris-Saclay, DSM/Irfu, CNRS UMR 7158, Université Paris Diderot, CEA Saclay, F-91191 Gif-sur-Yvette, France

**Abstract**

Zero-point quantum fluctuations of the electromagnetic vacuum create the widely known London-van der Waals attractive force between two atoms. Recently, there was a revived interest in the interaction of rotating matter with the quantum vacuum. Here, we consider a rotating pair of atoms maintained by London van der Waals forces and calculate the frictional torque they experience due to zero-point radiation. Using a semi-classical framework derived from the Fluctuation Dissipation Theorem, we take into account the full electrostatic coupling between induced dipoles. Considering the case of zero temperature only, we find a braking torque proportional to the angular velocity and to the third power of the fine structure constant. Although very small compared to London van der Waals attraction, the torque is strong enough to induce the formation of dimers in binary collisions. This new friction phenomenon at the atomic level should induce a paradigm change in the explanation of irreversibility.

We would like to answer a still vivid question: "What are the various routes of energy conversion between matter in relative motion and propagating electromagnetic [EM] fields?" Scientists have an extensive knowledge on the interactions between internal energy levels of atomic matter and EM waves[1]. However, it has long been thought that non-ionized gases (atomic or molecular) have negligible interactions with radiation. Furthermore Boltzmann's equation (the cornerstone of 20$^{th}$ century macroscopic physics) relies on the hypothetical ground of elastic, energy conserving atomic collisions[2]. Later, Collision Induced Absorption and Emission[3] studies, especially in atomic mixtures, evidenced the non-universality of Boltzmann's assumption. Meanwhile, the particular case of a pure atomic gas was long discarded due to a lack of quantitatively relevant interactions[3]. In contrast to this standard assumption, we find that the quantum vacuum exerts a non-negligible braking torque on two atoms rotating one around the other (equation (9) below).

Atomic matter in its ground state is neutral and does not carry any permanent dipolar or higher order electrostatic moment. However, London[4] showed with quantum mechanics that the fluctuating electrostatic atomic dipoles give way to the universal van der Waals forces discovered decades before. Forces induced by quantum fluctuations were later shown to be affected by EM propagation over long distances[5] (i.e. greater than atomic wavelengths) and to be macroscopically detectable[6]. After accurately measuring this Casimir force[7], research turned to the dynamical Casimir effect, also called Casimir friction, viz. on EM radiation emitted by macroscopic neutral bodies in relative motion[8].

Efforts were directed towards the calculation of dissipative components[9,10,11,12] of van der Waals forces between macroscopic bodies. The question of emissive collisions between atoms was also raised: are these forces conservative[13]? In this research, translational motion was mainly considered, with a general conclusion that the dissipated energy was negligible[14,15], especially in the zero-temperature case. On the contrary, we focus on rotational motion and find at zero-temperature a non-negligible friction.

Zel'dovich[16] suggests that a rotating body could amplify an incoming EM radiation, thus losing its rotational energy and angular momentum at the expense of the field. The specific case of the scattering of zero-point EM field by matter in rotation was studied decades later[17,18,19]. These authors treated the case of a rotating macroscopic body (dielectric or metallic) interacting with an EM field of variable temperature. Although nanoscopic materials could be considered, their EM characteristics were always represented by a dielectric constant, describing linear first order interaction within the material. The same is true with the usual treatment of fluctuation-induced interaction between atoms: the equations mainly deal with the perturbative term. Only recently, the self-consistent non-perturbative coupling was evoked[20].

Our approach combines both ways and considers two identical, neutral atoms rotating one around the other. The oscillator's dipoles and the vacuum field[21] are fluctuating quantum quantities. Their combined dynamics is treated herein via a semi classical framework derived from the Fluctuation Dissipation Theorem[22] [FDT]. A similar approach was taken previously[18] and shown to be an alternative to quantum treatments[23]. Research herein concentrates on zero temperature field only for which Milonni[21] thoroughly discussed the fluctuation-dissipation relation linking the vacuum field and an atomic dipole, already noticed by Callen and Welton[22].

In the frequency space, the FDT connects the self-correlations of a fluctuating physical quantity to the imaginary part of its response function. At zero temperature, the FDT for the components of the field at one location simply yields the self-correlation relation of the quantum vacuum:

$$\langle E_n[\omega] E_m[\omega'] \rangle = \frac{1}{3}\delta[\omega+\omega']\delta_{nm}\frac{\hbar|\omega|^3}{4\pi\varepsilon_o \pi c^3} \tag{1}$$

where the brackets $\langle ... \rangle$ represent the average over the fluctuations; $\delta[...]$ is Dirac's function and $\delta_{nm}$ is the Kronecker symbol. Herein two atoms interacting with the vacuum field are put into a non-equilibrium situation: mutual attraction and symmetry-breaking rotation. The FDT permits calculating the non-equilibrium behavior of the system from its equilibrium fluctuations. We will be using equation (1) to obtain the torque exerted by the field on the rotating pair of atoms. Before that, we need to express the polarizability of two rotating atoms.

A harmonic oscillator of natural angular frequency $\omega_o$, damped by radiation reaction describes an atom in its ground state. Applying[24,25] Newton's second law in the instantaneous inertial frame of the oscillator gives Abraham-Lorentz equation, which is Fourier transformed to obtain the atom's polarizability $\alpha[\omega]$:

$$\alpha[\omega] = \frac{q^2}{\mu}\frac{1}{\omega_o^2 - \omega^2 - i\tau\omega^3} \tag{2}$$

where $q$ is the charge in S.I. units, $\mu$ the reduced mass of the electron-nucleus system and $\tau$ the radiation reaction time[24,26] defined by $\tau = \frac{2(q^2/4\pi\varepsilon_o)}{3\mu c^3}$, where $c$ is the speed of light in vacuum. The polarizability, equation (2), relates the Fourier component of an atomic dipole to that of the local field that it experiences in an inertial frame. A long standing research[27,28] has tackled the well-known pathologies of Abraham-Lorentz equation, related to the impossibility of a point electron in classical physics. Corrections to Abraham-Lorentz approximation were developed in order to obtain a viable classical equation[28]. Contrary to this point of view, our semi-classical calculation treats dipoles immersed into a fluctuating vacuum field. This method allows approaching numerous properties of atoms interacting with the quantum vacuum[21]. Importantly, the FDT (1) is obtained[22] considering equilibrium between a fluctuating field and atomic dipoles the polarizability of which follow equation (2).

Now, two atoms rotate around their common center of mass. The distance separating them is $r$ and their angular velocity is $\Omega$ which is very small compared to $\omega_o$. Applying Newton's law to each atom in the inertial laboratory frame (in upper case letters in Fig. 1) yields two relations:

$$\ddot{\vec{P}}_1 - \tau\dddot{\vec{P}}_1 + \omega_o^2\vec{P}_1 = \frac{q^2}{\mu}\left(\vec{E}_1 + \frac{1}{4\pi\varepsilon_o}\frac{(3\vec{i}(\vec{i}.\vec{P}_2) - \vec{P}_2)}{r^3}\right) \tag{3-1}$$

$$\ddot{\vec{P}}_2 - \tau\dddot{\vec{P}}_2 + \omega_o^2\vec{P}_2 = \frac{q^2}{\mu}\left(\vec{E}_2 + \frac{1}{4\pi\varepsilon_o}\frac{(3\vec{i}(\vec{i}.\vec{P}_1) - \vec{P}_1)}{r^3}\right) \tag{3-2}$$

Each atom, *j*, feels a total field which is the sum of the fluctuating zero-point vacuum field, $\vec{E}_j$, at its position and the electrostatic dipolar field caused by the companion dipole. These equations include the well-known London van der Waals attraction[20], whether the atoms rotate or not. In the following, the attraction causes the centripetal acceleration. Distances between atoms remains short (i.e. $\omega_o r/c \ll 1$) in order to neglect the propagation of EM fields and to simplify calculations. Yet, $r$ remains significantly greater than $a_o$ (the Bohr radius) in order to neglect atomic repulsion. This hypothesis implies $\vec{E}_1 = \vec{E}_2 \equiv \vec{E}$.

Adding equations (3-1) and (3-2) gives the total dipole $\vec{P} = \vec{P}_1 + \vec{P}_2$ which obeys:

$$\ddot{\vec{P}} - \tau \dddot{\vec{P}} + \omega_o^2 \vec{P} = \frac{2q^2}{\mu}\vec{E} + \frac{q^2}{4\pi\varepsilon_o \mu r^3}\left(3(\cos\theta \vec{I} + \sin\theta \vec{J})(P_x\cos\theta + P_y\sin\theta) - (P_x\vec{I} + P_y\vec{J} + P_z\vec{K})\right) \quad (4)$$

After Fourier transform of equation (4) and some tedious algebra (cf. ref. 24 §2), the components of $\vec{P}$ result:

$$P_x[\omega] = \alpha_{xy}[\omega]\left(\frac{G_+[\omega] + \frac{3}{2}\frac{\alpha_{xy}[\omega+2\Omega]}{4\pi\varepsilon_o r^3}G_-[\omega+2\Omega]}{1 - \frac{9}{4}\frac{\alpha_{xy}[\omega+2\Omega]\alpha_{xy}[\omega]}{(4\pi\varepsilon_o r^3)^2}} + \frac{G_-[\omega] + \frac{3}{2}\frac{\alpha_{xy}[\omega-2\Omega]}{4\pi\varepsilon_o r^3}G_+[\omega-2\Omega]}{1 - \frac{9}{4}\frac{\alpha_{xy}[\omega-2\Omega]\alpha_{xy}[\omega]}{(4\pi\varepsilon_o r^3)^2}}\right) \quad (5\text{-}1)$$

$$P_y[\omega] = -i\alpha_{xy}[\omega]\left(\frac{G_+[\omega] + \frac{3}{2}\frac{\alpha_{xy}[\omega+2\Omega]}{4\pi\varepsilon_o r^3}G_-[\omega+2\Omega]}{1 - \frac{9}{4}\frac{\alpha_{xy}[\omega+2\Omega]\alpha_{xy}[\omega]}{(4\pi\varepsilon_o r^3)^2}} - \frac{G_-[\omega] + \frac{3}{2}\frac{\alpha_{xy}[\omega-2\Omega]}{4\pi\varepsilon_o r^3}G_+[\omega-2\Omega]}{1 - \frac{9}{4}\frac{\alpha_{xy}[\omega-2\Omega]\alpha_{xy}[\omega]}{(4\pi\varepsilon_o r^3)^2}}\right) \quad (5\text{-}2)$$

$$P_z[\omega] = 2\frac{q^2}{\mu}\frac{1}{\omega_z^2 - \omega^2 - i\tau\omega^3}E_z[\omega] \quad (5\text{-}3)$$

defining $\quad G_+[\omega] = E_x[\omega] + iE_y[\omega], \quad G_-[\omega] = E_x[\omega] - iE_y[\omega], \quad \alpha_{xy}[\omega] = \frac{q^2}{\mu(\omega_{xy}^2 - \omega^2 - i\tau\omega^3)},$

$\omega_{xy}^2 = \omega_o^2\left(1 - \frac{q^2}{2\mu\omega_o^2 4\pi\varepsilon_o r^3}\right) = \omega_o^2\left(1 - \frac{\alpha_o}{2r^3}\right)$ and $\omega_z^2 = \omega_o^2\left(1 + \frac{\alpha_o}{r^3}\right)$ with the usual definition $\alpha_o = \frac{q^2/4\pi\varepsilon_o}{\mu\omega_o^2}$.

By letting $\omega_o$ be the ionization frequency of the hydrogen atom and the volume $\alpha_o$ be $4a_o^3$, values of $r$ larger than $a_o$ result. For example, if $r > 5a_o$, then $\frac{\alpha_o}{r^3} \ll 1$ results. Equations (5-1) and (5-2) express the generalized susceptibility of $\vec{P}$, which mixes both field components at three different frequencies $(\omega, \omega \pm 2\Omega)$.

The total electric field exerts a torque on the oscillators' dipole, the fluctuation-averaged value of which $\vec{\Gamma}^{SC}$ is given by:

$$\vec{\Gamma}^{SC} = \langle \vec{P}[t] \wedge \vec{E}^{SC}[t]\rangle = \int_{\omega=-\infty}^{+\infty}\int_{\omega'=-\infty}^{+\infty}\langle \vec{P}[\omega] \wedge \vec{E}^{SC}[\omega']\rangle e^{-i\omega t}e^{-i\omega' t}d\omega d\omega' \quad (6)$$

where $\vec{E}^{SC}$ is the total [Self-Consistent] field seen by the atom, viz. the sum of the vacuum field and the induced field.

$$\vec{E}^{SC}[\omega'] = \frac{\vec{P}[\omega']}{2\alpha[\omega']} \quad (7)$$

We can now use[24] (5-1), (5-2), (7) and (1) to express the integrand of equation (6) as the sum of two integrals on all modes of the vacuum field:

$$\vec{\Gamma}^{SC} = \left(\Gamma_1^{SC} + \Gamma_2^{SC}\right)\vec{K} \quad (8\text{-}0)$$

$$\Gamma_1^{SC} = \int_{-\infty}^{+\infty} i\frac{|\alpha_{xy}[\omega]|^2}{\alpha^*[\omega]} \frac{2}{3} \frac{\hbar}{4\pi\varepsilon_o \pi c^3} |\omega|^3 \left( \frac{1}{\left|1 - \frac{9}{4}\frac{\alpha_{xy}[\omega+2\Omega]\alpha_{xy}[\omega]}{(4\pi\varepsilon_o r^3)^2}\right|^2} - \frac{1}{\left|1 - \frac{9}{4}\frac{\alpha_{xy}[\omega-2\Omega]\alpha_{xy}[\omega]}{(4\pi\varepsilon_o r^3)^2}\right|^2} \right) d\omega \quad (8\text{-}1)$$

$$\Gamma_2^{SC} = \int_{-\infty}^{+\infty} i\frac{|\alpha_{xy}[\omega]|^2}{\alpha^*[\omega]} \frac{2}{3} \frac{\hbar}{4\pi\varepsilon_o \pi c^3} \left( \frac{\left|\frac{3}{2}\frac{\alpha_{xy}[\omega+2\Omega]}{4\pi\varepsilon_o r^3}\right|^2 |\omega+2\Omega|^3}{\left|1 - \frac{9}{4}\frac{\alpha_{xy}[\omega+2\Omega]\alpha_{xy}[\omega]}{(4\pi\varepsilon_o r^3)^2}\right|^2} - \frac{\left|\frac{3}{2}\frac{\alpha_{xy}[\omega-2\Omega]}{4\pi\varepsilon_o r^3}\right|^2 |\omega-2\Omega|^3}{\left|1 - \frac{9}{4}\frac{\alpha_{xy}[\omega-2\Omega]\alpha_{xy}[\omega]}{(4\pi\varepsilon_o r^3)^2}\right|^2} \right) d\omega \quad (8\text{-}2)$$

where we have been using the isotropy of the vacuum field, with the * refering to the complex conjugate. $\Gamma_2^{SC}$ is a negligibly small quantity[24] which happens to be the only torque remaining in a first-order perturbative treatment of equations (3-1) and (3-2). $\Gamma_1^{SC}$ depends entirely on the self-consistent interaction.

Using a few changes of variable and developing in Taylor series (cf. ref 24 §5) with respect to the small parameters $\Omega/\omega_o$ and $\beta = \tau\omega_o$, one obtains:

$$\Gamma_1^{SC} \cong -\frac{9}{8}\tau\omega_o \left(\frac{\alpha_o}{r^3}\right)^2 \hbar\Omega \quad (9)$$

where the factor $\tau\omega_o$ equals nearly the third power of the fine structure constant[24], numerically about $10^{-7}$. Although equation (9) occurs through a Taylor expansion of the integrand of equation (8-1), its validity is much wider than the quality of the expansion could suggest. The numerical integration of (8-1) yields nearly exactly (9) for numerous tested values in the range $\frac{\Omega}{\omega_o} \leq 10^{-3}$ and $\frac{\alpha_o}{r^3} \ll 10^{-1}$. The torque (9) was obtained as the effect of the total (vacuum + induced) field on the self-consistent dipoles. The same result occurs when one considers the effect of the vacuum field only on those dipoles, within first order in $\frac{\Omega}{\omega_o}$.

The braking torque given by equation (9) is the main result of this communication. The equivalent tangential braking force is very small compared to the van der Waals attractive force, their ratio being of order $\tau\Omega$: for example $\tau\Omega \approx 10^{-11}$ if $\frac{\Omega}{\omega_o} \approx 10^{-4}$. Nevertheless, this torque decreases the kinetic energy and the angular momentum of the atoms. A linear torque gives rise to a temporally exponential attenuation of the angular momentum with a characteristic time $T$, depending on $r$:

$$T[r] = \frac{\frac{1}{2}Mr^2}{\frac{9}{8}\left(\frac{\alpha_o}{r^3}\right)^2 \tau \omega_o \hbar} \qquad (10)$$

with $M$ the mass of one atom. $T$ is on order $10^{-2}$ s for the numerical case already considered, with $r \approx 5a_o$ and M the mass of the hydrogen atom. Before discussing the physical results, two comments on the order of magnitude are warranted. On the one hand, the braking time given by equation (10) is rather long compared to the duration of most atomic collisions, generally on order $10^{-10}$ s or shorter, but it could be relevant when macroscopic processes are at stake. In particular, this quantum friction effect should be considered as a noticeable contribution to energy dissipation and entropy growth in gaseous systems. Within our theoretical development, the friction phenomenon shall be present in any atomic, or molecular, interaction. Thus the standard explanation of irreversibility should be revised: instead of resting on probability considerations, it could be derived from the universal existence of dynamical friction forces between atomic structures. This attractive task is nevertheless secondary compared to the experimental testing of equation (9), which is briefly discussed below.

On the other hand, the braking torque (9) might be relevant macroscopically, but it is small enough to have stayed unnoticed, and to have been overlooked in the past. Let us now compare this theoretical result with other published work, and discuss its consequences and testability.

As recalled above, Casimir friction at zero-temperature was previously considered for rectilinear motion mainly. In the case of a metallic plate sliding at a fixed distance from a second similar plate[11,29] with a relative velocity $v$, the friction force was found scaling with $v^3$. A cubic power law in velocity was also obtained recently[30,31] for the atom-surface drag force at zero temperature, in contrast to several different results previously published on this Casimir-Polder configuration. The material and geometric hypotheses of ref. 11 and 29-31 differ from the case of two rotating atoms for which we find a friction torque (9) linear in the azimuthal velocity $r\Omega$. The discrepancy is thus not surprising, but more work would be necessary to explain it. In a different perspective, also considering a pure translational motion, Boye and Brevik[14] and independently Barton[15] calculated the energy loss in an atomic collision. They neglected the effect of van der Waals attraction on the trajectory, consequently forbidding any rotation of the atoms. They found a very small friction at zero-temperature, which varied as $\exp[-v]$ and was totally negligible for non-relativistic velocities. By taking the effect of van der Waals forces into account and integrating the full electrostatic coupling we find a very different result, linear in $r\Omega$.

Previous calculations also considered rotating media[18,19]. They found the effect of vacuum friction to be negligible on isolated dielectrics in rotation. The interaction between the atoms in those materials was considered to be at equilibrium, giving rise to a polarizability, or dispersion relation, insensitive to thermal and mechanical parameters. Herein, on the contrary, the dissipative torque results from the strong dependence of the two atoms' polarizability on the interatomic distance and from the self-consistent treatment of their interactions. It is fair to note that a different polarizability function would result in a different velocity dependence. In another configuration, a conductive sphere rotating near a surface[32] experiences a frictional torque scaling as $(r\Omega)^3$ at zero temperature and as $r\Omega$ at high temperatures.

Further work is needed to give a comprehensive description of quantum friction in all these diverse configurations where the physics of momentum transfer is similar: virtual photons are exchanged by the atoms, resulting on average in momentum loss by the material system. A related concern of former work is of interest to our result. According to ref. 19, a rotating body would drag along nearby objects and share its angular momentum with them, through the vacuum field. Herein, the question is, "How and how much can a rotating pair of atoms influence the motion of another pair in the vicinity?" This drives the attention to the physical ways by which the energy is radiated away.

The radiation reaction term leads to energy and angular momentum loss, and it involves the emission of an outgoing wave. This radiation takes place via photon emission, which cannot be tackled within the present semi-classical framework. The emission phenomenon should be the subject of further work. Nevertheless, we can still deduce two properties of this EM emission. First, it is characterized by its energy and angular momentum outflow, the ratio of which is the average frequency, $\Omega$. Second, due to the symmetry of the system the emission process shall not carry any linear momentum.

Apart from experimental tests by detection of EM emission, the consequences of (9) should be studied in the mechanism for dimer formation[33]. With strictly conservative interactions, a third body is needed to induce the capture of one atom by another. Figure 2 illustrates how equation (9) can change the situation of a binary collision. Due to the $1/r^6$ attractive potential, the two atoms will classically experience a so-called "centrifugal barrier" of height depending on their angular momentum (cf. ref. 33 § 4.2). At or near the barrier, the two atoms orbit extensively, thus resembling the case of Fig. 1. Slowed down by (9), the atoms can "fall" into one of the bound dimer states.

A barrier towards the experimental test of the torque (9) is its smallness, due in part to the atomic polarizability $\alpha_o$ of order the atomic volume $a_o^3$. However the effective size of the atom depends on its excitation level. For example, very recent experiments[34] provided the first direct measurement of the attractive van der Waals force between atoms in Rydberg states, the size of which is much larger than $a_o$. The same kind of systems could be used to detect a potential dissipative torque.

The vacuum friction expressed in equation (9) is strong enough to induce a paradigm change in the explanation of irreversibility. But any attempt to reach such a goal should be aware of two other pending jobs. On the one hand, experimental testing awaits the design of dedicated experiments. On the other hand, further steps on the theoretical side should include the extension of this semi-classical calculation with the tools of quantum electrodynamics.

**Acknowledgements**

We benefited from interesting discussions with F. Ladieu, F. Daviaud, Y. Pomeau, M. Le Berre, H. Desvaux, L. Desvillettes, B. Dubrulle, G. Ferrando, F. Graner, A. Lambrecht, K. Mallick, J.-M. Mestdagh, S. Reynaud and C. Rountree. We very much appreciated a helpful reading and calculation by an anonymous referee. CEA provided its support through DSM-Énergie programme.

**FIGURES**

**Figure 1: A rotating pair of atoms**

Atomic pair represented in its rotation plane. In the inertial frame $\vec{I}, \vec{J}, \vec{K}$ (upper case letters), the segment linking the two oscillators turns with angular velocity $\Omega$; its unit vector, fixed in the rotating frame $\vec{i}, \vec{j}, \vec{k}$ (italic lower case letters), is $\vec{i} = \cos\theta \vec{I} + \sin\theta \vec{J}$ with $\theta = \Omega t$.

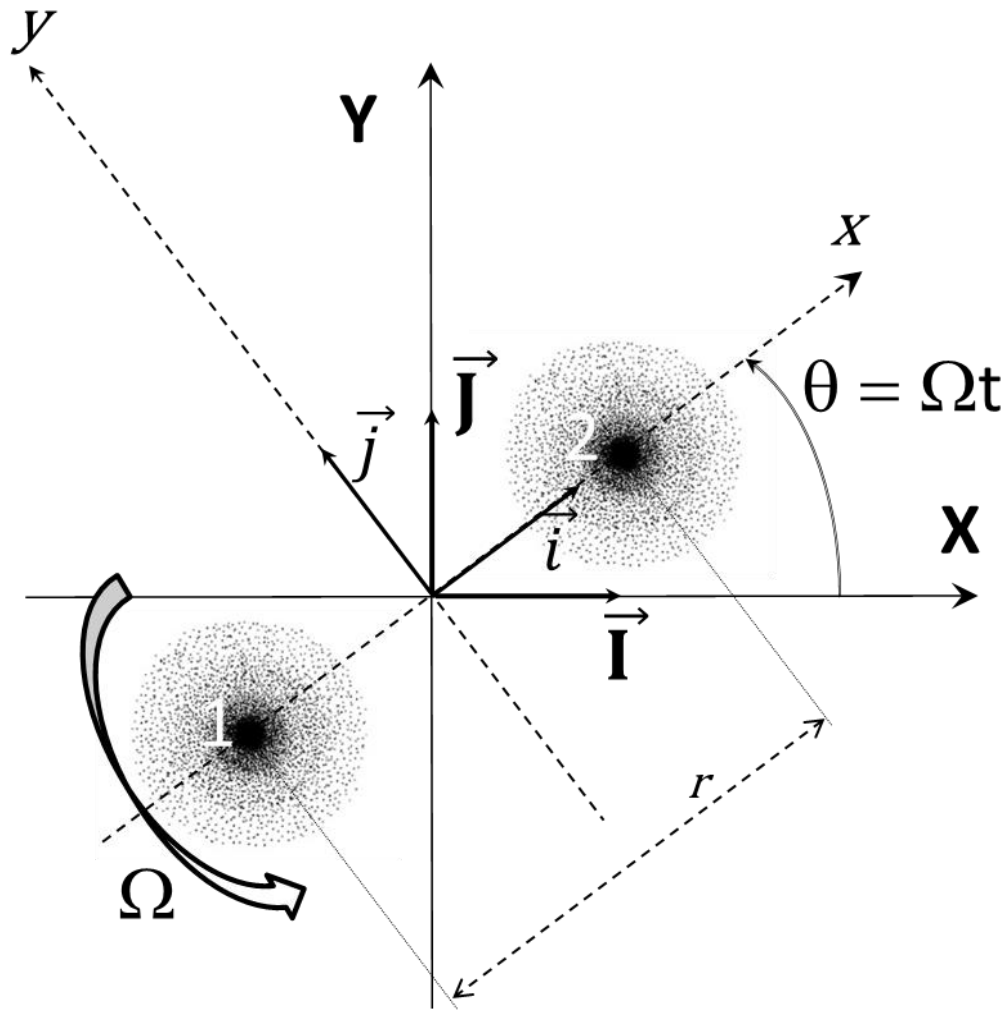

**Figure 2: Effect of the dissipative torque on a binary collision**

a) The effective potential (van der Waals + centrifugal) seen by the two-atom system depends on the angular momentum $L$. Effective energy for $L^2 = 10\hbar^2$ (solid line); the apex is at $\dfrac{r}{a_o} \sim 4.5$ for an effective energy $V_{\text{eff}}$ (dashed line).

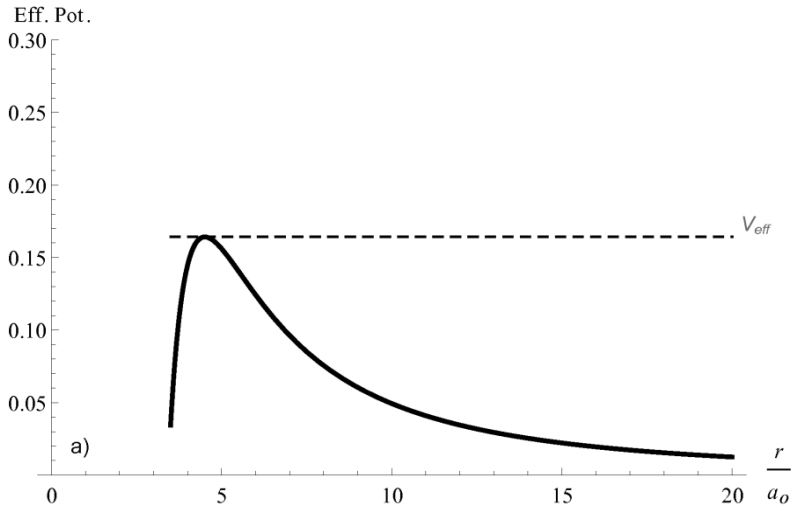

b) Numerical integration of the classical dynamics of the two atoms with $L^2 = 10\hbar^2$ and effective energy slightly under $V_{\text{eff}}$. The plain line takes into account the dissipative torque (9) while the dashed line does not. In both cases the system is "orbiting" at $\dfrac{r}{a_o} \sim 4.5$. If no energy is lost to the field, the dynamics is reversible and the atoms finally separate. Vacuum friction prevents the separation, and finally induces the two atoms to fall one towards the other.

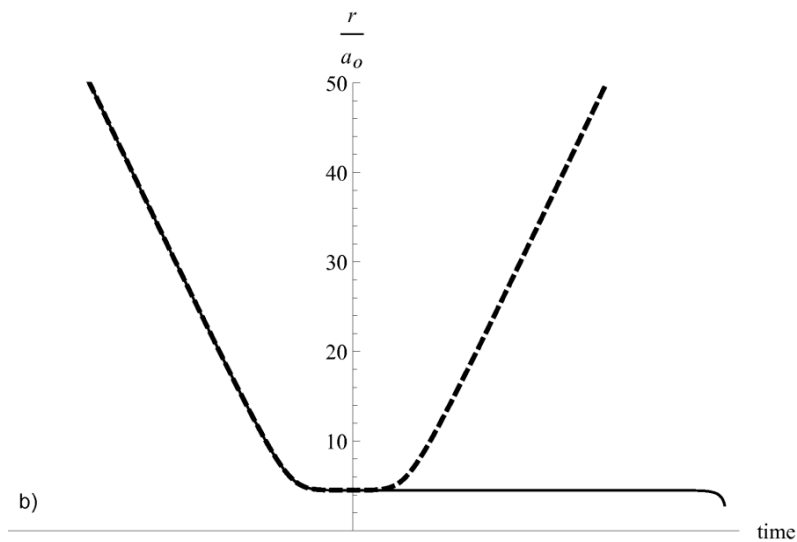

SUPPLEMENTAL MATERIAL

We are going to expose various lengthy or tricky developments necessary to obtain the results presented in the main text. Here are the topics that will be successively treated:

## 1. An atomic oscillator in an external field

Applying Newton's second law in the instantaneous inertial frame of the oscillator, we obtain Abraham-Lorentz equation[1,2] for the atomic dipole $\overrightarrow{P_a}$ in an external field $\overrightarrow{E_{ext}}$:

$$\ddot{\overrightarrow{P_a}} - \tau \dddot{\overrightarrow{P_a}} + \omega_o^2 \overrightarrow{P_a} = \frac{q^2}{\mu} \overrightarrow{E_{ext}} \quad \text{(A-1)}$$

where q is the charge in S.I. units, $\mu$ the reduced mass of the electron-proton system and $\tau$ the radiation reaction time defined by $\tau = \frac{2(q^2/4\pi\varepsilon_o)}{3\mu c^3}$, with c the speed of light in vacuum ("S" in the equation numbering stands for "Supplement"; numbers without S refer to the main text). For numerical estimates, we will take for $\omega_o$ the ionization angular frequency of Hydrogen, $\omega_o = \frac{(q^2/4\pi\varepsilon_o)^2 \mu}{2\hbar^3} \approx 2.1 \; 10^{16} \text{ rad s}^{-1}$.

Because the radiation reaction damping term contains a third derivative of particles' position (in the dipole moment), it has been criticized for introducing non causal behaviors and runaway solutions. However, Feynman[3] explained clearly the occurrence of the radiation reaction term, while Milonni[21] discussed its indissoluble link with the vacuum energy density varying as $\omega^3$ (see ref. 3 chapter 28, especially §28-3, and ref. 21 chapter 5). The two cited pathologies do not concern bound and periodic trajectories, and thus can be ignored here.

Equation (A-1) neglects the effect of the magnetic field, a standard approximation when considering non relativistic motion. The atomic polarizability $\alpha[\omega]$ is obtained by Fourier transforming (A-1)

$$\alpha[\omega] = \frac{q^2}{\mu} \frac{1}{\omega_o^2 - \omega^2 - i\tau\omega^3} \quad \text{(A-2)}$$

where we took the convention that square brackets […] contain the argument of a function. In the following, we will systematically use (A-2) to relate the Fourier component of an atomic dipole and that of the local field that it experiences in an inertial frame. When we study products of two real quantities, for example in average energy calculations, we finally manipulate products of components of the field:

---

[1] P. W. Milonni, *The quantum vacuum: an introduction to quantum electrodynamics* (Academic press, San Diego, 1994).
[2] J. D. Jackson, *Classical Electrodynamics* (Wiley, New York, 1999).
[3] R. P. Feynman, R. B. Leighton and M. Sands, *The Feynman Lectures on Physics, Vol. 2: Mainly Electromagnetism and Matter* (Addison-Wesley, Reading, 1964).

$$\langle \vec{P}[t].\vec{E}[t]\rangle = \int_{-\infty}^{\infty}\int_{-\infty}^{\infty} \alpha[\omega]\vec{E}[\omega].\vec{E}[\omega']e^{-i\omega t}e^{-i\omega' t}d\omega\,d\omega' \tag{A-3}$$

Applying the Fluctuation Dissipation Theorem, equation (1) in the main text, we are left with products of components at ω and at ω'=-ω. The fact that $\vec{E}[t]$ is real induces that $\vec{E}[-\omega] = (\vec{E}[\omega])^*$ where the star stands for the complex-conjugate; thus the integral (A-3) reduces to squared moduli of the field components.

## 2. Dipole moments of the rotating atoms

In order to obtain the polarizability of the two atoms, we apply Newton's law for each atom in the inertial laboratory frame (in upper case letters in Fig. 1 in the main text). We first get a relation on $\vec{P} = \vec{P}_1 + \vec{P}_2$ :

$$\ddot{\vec{P}} - \tau\dddot{\vec{P}} + \omega_o^2\vec{P} = \frac{2q^2}{\mu}\left(E_x\vec{I} + E_y\vec{J} + E_z\vec{K}\right) + \frac{e^2}{\mu r^3}\left(3(\cos\theta\vec{I} + \sin\theta\vec{J})(P_x\cos\theta + P_y\sin\theta) - (P_x\vec{I} + P_y\vec{J} + P_z\vec{K})\right) \tag{A-4}$$

where $\vec{E} = \frac{\vec{E}_1+\vec{E}_2}{2}$ is the average field between atoms 1 and 2. As discussed in the main text, we neglect the propagation of EM fields. Thus, in the following $\vec{E}_1$ and $\vec{E}_2$ are equal to $\vec{E}$. Equation (A-4) is further developed and projected on the three axes:

$$\ddot{P}_x - \tau\dddot{P}_x + \omega_o^2 P_x = \frac{2q^2}{\mu}E_x + \frac{e^2}{\mu r^3}\left(\left(\frac{3}{2}\cos 2\theta + \frac{1}{2}\right)P_x + \frac{3}{2}\sin 2\theta P_y\right) \tag{A-5-x}$$

$$\ddot{P}_y - \tau\dddot{P}_y + \omega_o^2 P_y = \frac{2q^2}{\mu}E_y + \frac{e^2}{\mu r^3}\left(\frac{3}{2}\sin 2\theta P_x + \left(\frac{1}{2} - \frac{3}{2}\cos 2\theta\right)P_y\right) \tag{A-5-y}$$

$$\ddot{P}_z - \tau\dddot{P}_z + \omega_o^2 P_z = \frac{2q^2}{\mu}E_z - \frac{e^2}{\mu r^3}P_z \tag{A-5-z}$$

After Fourier transforming, (A-5-z) yields :

$$P_z[\omega] = 2\frac{q^2}{\mu}\frac{1}{\omega_z^2 - \omega^2 - i\tau\omega^3}E_z = \alpha_z[\omega]E_z \tag{A-6-z}$$

with $\omega_z^2 = \omega_o^2\left[1 + \frac{e^2}{\mu\omega_o^2 r^3}\right] = \omega_o^2\left[1 + \frac{\alpha_o}{r^3}\right]$, and the usual definition $\alpha_o = \frac{q^2/4\pi\varepsilon_o}{\mu\omega_o^2}$. $\omega_o$ being the ionization frequency of the Hydrogen atom, the volume $\alpha_o$ equals $4a_o^3$, with $a_o$ the Bohr radius. We consider values of $r$ larger than $a_o$, say $r \geq 5\,a_o$, so that $\frac{\alpha_o}{r^3} \ll 1$. Equations (A-5-x) and (A-5-y) reveal the important feature of the rotating system: the two field components in the rotation plane are mixed. This mixing can be further worked out by Fourier transforming the equations. That for, we need to pay attention to the cos2θ and sin2θ terms. Being time dependent, their Fourier transforms introduce the angular frequency Ω of the rotating frame:

$$-\omega^2 P_x[\omega] - i\tau\omega^3 P_x[\omega] + \omega_o^2 P_x[\omega] = \frac{2q^2}{\mu}E_x[\omega] + \frac{e^2}{\mu r^3}\left(\frac{3}{2}\frac{P_x[\omega+2\Omega]+P_x[\omega-2\Omega]}{2} + \frac{P_x[\omega]}{2} + \frac{3}{2}\frac{P_y[\omega+2\Omega]-P_y[\omega-2\Omega]}{2i}\right) \tag{A-6-x}$$

$$-\omega^2 P_y[\omega] - i\tau\omega^3 P_y[\omega] + \omega_o^2 P_y[\omega] = \frac{2q^2}{\mu}E_y[\omega] + \frac{e^2}{\mu r^3}\left(\frac{3}{2}\frac{P_x[\omega+2\Omega]-P_x[\omega-2\Omega]}{2i} + \frac{P_y[\omega]}{2} - \frac{3}{2}\frac{P_y[\omega+2\Omega]+P_y[\omega-2\Omega]}{2}\right) \tag{A-6-y}$$

In order to decouple $P_x$ and $P_y$ and to express $\vec{P}$ as a function of field modes, we introduce 2 new dipole functions

$$F_+[\omega] = P_x[\omega] + iP_y[\omega] \qquad F_-(\omega) = P_x[\omega] - iP_y[\omega] \tag{A-7}$$

and two new field functions

$$G_+[\omega] = E_x[\omega] + iE_y[\omega] \qquad G_-[\omega] = E_x[\omega] - iE_y[\omega] \qquad (A-8)$$

We add (A-6-x) and $i$ times (A-6-y) to obtain (A-9)

$$-\omega^2 F_+[\omega] - i\tau\omega^3 F_+[\omega] + \left(\omega_o^2 - \frac{e^2}{2\mu r^3}\right) F_+[\omega] = \frac{2q^2}{\mu} G_+[\omega] + \frac{3}{2}\frac{e^2}{\mu r^3} F_-[\omega + 2\Omega] \qquad (A-9)$$

Similarly, subtracting $i$ times (A-6-y) from (A-6-x) leads to (A-10)

$$-\omega^2 F_-[\omega] - i\tau\omega^3 F_-[\omega] + \left(\omega_o^2 - \frac{e^2}{2\mu r^3}\right) F_-[\omega] = \frac{2q^2}{\mu} G_-[\omega] + \frac{3}{2}\frac{e^2}{\mu r^3} F_+[\omega - 2\Omega] \qquad (A-10)$$

We will now write $\omega_{xy}^2 = \omega_o^2\left[1 - \frac{e^2}{2\mu\omega_o^2 r^3}\right] = \omega_o^2\left[1 - \frac{\alpha_o}{2r^3}\right]$ and $\alpha_{xy}[\omega] = \frac{q^2}{\mu(\omega_{xy}^2 - \omega^2 - i\tau\omega^3)}$. Using the simple trick of applying (A-10) at $\omega + 2\Omega$, we get:

$$F_-[\omega + 2\Omega] = \left\{2\alpha_{xy}[\omega + 2\Omega]G_-[\omega + 2\Omega] + \frac{1}{4\pi\varepsilon_o r^3}\frac{3}{2}\alpha_{xy}[\omega + 2\Omega]F_+[\omega]\right\}$$

which, injected into (A-9), leads to

$$(\omega_{xy}^2 - \omega^2 - i\tau\omega^3)F_+[\omega] = \frac{2q^2}{\mu} G_+[\omega] + \frac{3}{2}\frac{\alpha_{xy}[\omega+2\Omega]}{4\pi\varepsilon_o r^3}\left\{\frac{2q^2}{\mu} G_-[\omega + 2\Omega] + \frac{e^2}{\mu r^3}\frac{3}{2}F_+[\omega]\right\}$$

transformed into

$$F_+[\omega] = 2\alpha_{xy}[\omega] \frac{G_+[\omega] + \frac{3}{2}\frac{\alpha_{xy}[\omega+2\Omega]}{4\pi\varepsilon_o r^3}G_-[\omega+2\Omega]}{1 - \frac{9\alpha_{xy}[\omega+2\Omega]\alpha_{xy}[\omega]}{4}\frac{1}{(4\pi\varepsilon_o r^3)^2}} \qquad (A-11)$$

Another similar trick is applied by taking (A-9) at $\omega - 2\Omega$

$$F_+[\omega - 2\Omega] = 2\alpha_{xy}[\omega - 2\Omega]G_+[\omega - 2\Omega] + \frac{3}{2}\frac{\alpha_{xy}[\omega-2\Omega]}{4\pi\varepsilon_o r^3} F_-[\omega]$$

a relation that we inject into (A-10) to obtain

$$(\omega_{xy}^2 - \omega^2 - i\tau\omega^3)F_-[\omega] = \frac{2q^2}{\mu} G_-[\omega] + \frac{3}{2}\frac{e^2}{\mu r^3}\left(2\alpha_{xy}[\omega - 2\Omega]G_+[\omega - 2\Omega] + \frac{3}{2}\frac{\alpha_{xy}[\omega-2\Omega]}{4\pi\varepsilon_o r^3}F_-[\omega]\right)$$

transformed into

$$F_-[\omega] = 2\alpha_{xy}[\omega] \frac{G_-[\omega] + \frac{3}{2}\frac{\alpha_{xy}[\omega-2\Omega]}{4\pi\varepsilon_o r^3}G_+[\omega-2\Omega]}{1 - \frac{9\alpha_{xy}[\omega-2\Omega]\alpha_{xy}[\omega]}{4}\frac{1}{(4\pi\varepsilon_o r^3)^2}} \qquad (A-12)$$

Using (A-7), we easily obtain the components of $\vec{P}$ from (A-11) and (A-12):

$$P_x[\omega] = \alpha_{xy}[\omega]\left(\frac{G_+[\omega] + \frac{3}{2}\frac{\alpha_{xy}[\omega+2\Omega]}{4\pi\varepsilon_o r^3}G_-[\omega+2\Omega]}{1 - \frac{9\alpha_{xy}[\omega+2\Omega]\alpha_{xy}[\omega]}{4}\frac{1}{(4\pi\varepsilon_o r^3)^2}} + \frac{G_-[\omega] + \frac{3}{2}\frac{\alpha_{xy}[\omega-2\Omega]}{4\pi\varepsilon_o r^3}G_+[\omega-2\Omega]}{1 - \frac{9\alpha_{xy}[\omega-2\Omega]\alpha_{xy}[\omega]}{4}\frac{1}{(4\pi\varepsilon_o r^3)^2}}\right) \qquad (A-13-x)$$

$$P_y[\omega] = -i\alpha_{xy}[\omega]\left(\frac{G_+[\omega] + \frac{3}{2}\frac{\alpha_{xy}[\omega+2\Omega]}{4\pi\varepsilon_o r^3}G_-[\omega+2\Omega]}{1 - \frac{9\alpha_{xy}[\omega+2\Omega]\alpha_{xy}[\omega]}{4}\frac{1}{(4\pi\varepsilon_o r^3)^2}} - \frac{G_-[\omega] + \frac{3}{2}\frac{\alpha_{xy}[\omega-2\Omega]}{4\pi\varepsilon_o r^3}G_+[\omega-2\Omega]}{1 - \frac{9\alpha_{xy}[\omega-2\Omega]\alpha_{xy}[\omega]}{4}\frac{1}{(4\pi\varepsilon_o r^3)^2}}\right) \qquad (A-13-y)$$

## 3. Torque exerted by the vacuum field

We are interested in the fluctuation-averaged value of the self-consistent (SC) torque $\vec{\Gamma}^{SC}$

$$\vec{\Gamma}^{SC} = \langle \vec{P}_1[t] \wedge \vec{E}^{SC}{}_1[t] + \vec{P}_2[t] \wedge \vec{E}^{SC}{}_2[t] \rangle$$

where $\vec{E}^{SC}{}_n$ is the total field seen by atom n. The hypothesis $\vec{E}_1 = \vec{E}_2$ yields directly $\vec{P}_1 = \vec{P}_2 = \frac{\vec{P}}{2}$. In the frequency space, the total field is then $\vec{E}^{SC}[\omega] = \frac{\vec{P}[\omega]}{2\alpha[\omega]}$. We thus obtain the self-consistent torque through

$$\vec{\Gamma}^{SC} = \int_{\omega=-\infty}^{+\infty}\int_{\omega'=-\infty}^{+\infty} 2\langle \frac{\vec{P}}{2}[\omega] \wedge \frac{\vec{P}[\omega']}{2\alpha[\omega']} \rangle e^{-i\omega t}e^{-i\omega' t}d\omega d\omega' \tag{A-14}$$

The integrand of (A-14) is rewritten

$$\frac{1}{2}\langle \vec{P}[\omega] \wedge \frac{\vec{P}[\omega']}{\alpha[\omega']} \rangle = \frac{1}{2}\frac{1}{\alpha[\omega']}\langle P_x[\omega]P_y[\omega'] - P_x[\omega']P_y[\omega]\rangle \vec{K} \equiv \gamma^{SC}[\omega,\omega',\Omega]\vec{K} \tag{A-15}$$

and is worked out in order to apply the FDT relation of the field (1). In (A-15), the cross products involving the z components have been discarded, because according to (1) they bring null terms only. Using equations (5) ((A-13-x) and (A-13-y) in the Supplemental Material), we obtain a rather heavy relation mixing products of $G_\pm[\omega]$ and $G_\pm[\omega']$. From (A-8) and (1), we deduce the FDT relations for $G_\pm$.

$$\langle G_+[\omega]G_+[\omega']\rangle = \langle G_-[\omega]G_-[\omega']\rangle = \frac{1}{3}(\delta[\omega+\omega'] - \delta[\omega+\omega'])\frac{\hbar|\omega|^3}{4\pi\varepsilon_o\pi c^3} = 0$$

$$\langle G_+[\omega]G_-[\omega']\rangle = \langle G_-[\omega']G_+[\omega]\rangle = \frac{1}{3}(\delta[\omega+\omega'] + \delta[\omega+\omega'])\frac{\hbar|\omega|^3}{4\pi\varepsilon_o\pi c^3} = \frac{2}{3}\delta[\omega+\omega']\frac{\hbar|\omega|^3}{4\pi\varepsilon_o\pi c^3} \tag{A-16}$$

The relations (A-16) will select in the integral (A-14) the $\omega'$ that will give non-zero contributions. Since $G_\pm[\omega]$ contains field components at three different frequencies $(\omega,\omega\pm 2\Omega)$, we have to check the terms obtained with $\omega' \in \{-\omega, -\omega \mp 2\Omega\}$. Terms originated from $\omega' = -\omega \mp 2\Omega$ finally cancel out, as expected since they would have induced contributions to the torque varying with $\exp[\pm 2i\Omega t]$. The only non-zero contributions come from the case $\omega' = -\omega$:

$$\gamma^{SC}[\omega,-\omega,\Omega] = i\frac{|\alpha_{xy}[\omega]|^2}{\alpha^*[\omega]}\langle G_+[\omega]G_-[-\omega]\rangle \left( \frac{1}{\left|1 - \frac{9}{4}\frac{\alpha_{xy}[\omega+2\Omega]\alpha_{xy}[\omega]}{(4\pi\varepsilon_o r^3)^2}\right|^2} - \frac{1}{\left|1 - \frac{9}{4}\frac{\alpha_{xy}[\omega-2\Omega]\alpha_{xy}[\omega]}{(4\pi\varepsilon_o r^3)^2}\right|^2} \right) +$$

$$i\frac{|\alpha_{xy}[\omega]|^2}{\alpha^*[\omega]}\langle G_+[\omega]G_-[-\omega]\rangle \left( \frac{\left|\frac{3}{2}\frac{\alpha_{xy}[\omega+2\Omega]}{4\pi\varepsilon_o r^3}\right|^2 \langle G_-[\omega+2\Omega]G_+[-\omega-2\Omega]\rangle}{\left|1 - \frac{9}{4}\frac{\alpha_{xy}[\omega+2\Omega]\alpha_{xy}[\omega]}{(4\pi\varepsilon_o r^3)^2}\right|^2} - \frac{\left|\frac{3}{2}\frac{\alpha_{xy}[\omega-2\Omega]}{4\pi\varepsilon_o r^3}\right|^2 \langle G_+[\omega-2\Omega]G_-[-\omega+2\Omega]\rangle}{\left|1 - \frac{9}{4}\frac{\alpha_{xy}[\omega-2\Omega]\alpha_{xy}[\omega]}{(4\pi\varepsilon_o r^3)^2}\right|^2} \right)$$

(A-17)

The integration of (A-17) will yield two contributions to the torque $\vec{\Gamma}^{SC} = \left(\Gamma_1^{SC} + \Gamma_2^{SC}\right)\vec{K}$ expressed in (8-1) and (8-2) of the main text. Using the property $\alpha[-\omega] = \alpha^*[\omega]$, these integrals are easily transformed into

$$\Gamma_1^{SC} = -2\frac{2}{3}\frac{\hbar}{4\pi\varepsilon_o \pi c^3}\int_0^{+\infty}\frac{\text{Im}[\alpha[\omega]]}{|\alpha[\omega]|^2}|\alpha_{xy}[\omega]|^2 |\omega|^3 \left(\frac{1}{\left|1-\frac{9}{4}\frac{\alpha_{xy}[\omega+2\Omega]\alpha_{xy}[\omega]}{(4\pi\varepsilon_o r^3)^2}\right|^2} - \frac{1}{\left|1-\frac{9}{4}\frac{\alpha_{xy}[\omega-2\Omega]\alpha_{xy}[\omega]}{(4\pi\varepsilon_o r^3)^2}\right|^2}\right)d\omega \quad \text{(A-18-1)}$$

$$\Gamma_2^{SC} = -3\frac{\hbar}{\pi c^3}\int_0^{+\infty}\frac{\text{Im}[\alpha[\omega]]}{|\alpha[\omega]|^2}|\alpha_{xy}[\omega]|^2 \left(\frac{|\omega+2\Omega|^3 \left|\frac{\alpha_{xy}[\omega+2\Omega]}{4\pi\varepsilon_o r^3}\right|^2}{\left|1-\frac{9}{4}\frac{\alpha_{xy}[\omega+2\Omega]\alpha_{xy}[\omega]}{(4\pi\varepsilon_o r^3)^2}\right|^2} - \frac{|\omega-2\Omega|^3 \left|\frac{\alpha_{xy}[\omega-2\Omega]}{4\pi\varepsilon_o r^3}\right|^2}{\left|1-\frac{9}{4}\frac{\alpha_{xy}[\omega-2\Omega]\alpha_{xy}[\omega]}{(4\pi\varepsilon_o r^3)^2}\right|^2}\right)d\omega \quad \text{(A-18-2)}$$

where one shall pay attention to the new integration range, from 0 to $+\infty$. We will now use the relation:

$$\frac{\text{Im}[\alpha[\omega]]}{|\alpha[\omega]|^2} = \frac{\mu}{q^2}\tau\omega^3 = \frac{2}{3}\frac{\omega^3}{4\pi\varepsilon_o c^3}$$

We begin with the evaluation of $\Gamma_2^{SC}$

$$\Gamma_2^{SC} = -2\frac{1}{c^3}\frac{\hbar}{\pi c^3}\left(\int_0^{+\infty}\omega^3|\omega+2\Omega|^3 \frac{\left|\frac{\alpha_{xy}[\omega]}{4\pi\varepsilon_o}\right|^2 \left|\frac{\alpha_{xy}[\omega+2\Omega]}{4\pi\varepsilon_o r^3}\right|^2}{\left|1-\frac{9}{4}\frac{\alpha_{xy}[\omega+2\Omega]\alpha_{xy}[\omega]}{(4\pi\varepsilon_o r^3)^2}\right|^2}d\omega - \int_0^{+\infty}\omega^3|\omega-2\Omega|^3 \frac{\left|\frac{\alpha_{xy}[\omega]}{4\pi\varepsilon_o}\right|^2 \left|\frac{\alpha_{xy}[\omega-2\Omega]}{4\pi\varepsilon_o r^3}\right|^2}{\left|1-\frac{9}{4}\frac{\alpha_{xy}[\omega-2\Omega]\alpha_{xy}[\omega]}{(4\pi\varepsilon_o r^3)^2}\right|^2}d\omega\right) \quad \text{(A-19)}$$

Symmetry properties of the integrand in (A-19) allow simplifying drastically the calculation by a change of variable $\omega \to \omega + 2\Omega$ in the first term. Finally,

$$\Gamma_2^{SC} = 2\frac{1}{c^3}\frac{\hbar}{\pi c^3}\int_0^{2\Omega}\omega^3|\omega-2\Omega|^3 \left|\frac{\alpha_{xy}[\omega]}{4\pi\varepsilon_o}\right|^2 \left|\frac{\alpha_{xy}[\omega-2\Omega]}{4\pi\varepsilon_o r^3}\right|^2 d\omega \quad \text{(A-20)}$$

The integral of (A-20) is limited to frequencies $\omega$ smaller than $2\Omega$. Since $\Omega \ll \omega_o$, we consider $\alpha_{xy}[\omega] \cong \alpha_{xy}[0]$ for $0 < \omega < 2\Omega$. The lowest order term in $\frac{\alpha_o}{r^3}$ is

$$\Gamma_2^{SC} \cong +(\tau\omega_o)^2 \frac{144}{35\pi}\left(\frac{\alpha_o}{r^3}\right)^2 \left(\frac{\Omega}{\omega_o}\right)^7 \hbar\omega_o \quad \text{(A-21)}$$

Two things need to be noticed about (A-21). First $\Gamma_2^{SC}$ is of the same sign as $\Omega$ thus it is an accelerating torque: physically, we expect the other part $\Gamma_1^{SC}$ to be of opposite sign, and to at least compensate for $\Gamma_2^{SC}$. Secondly, the dependence of (A-21) in $\Omega^7$ prevents it from playing any quantitative role in standard atomic dynamics. Taking $\Omega \approx 10^{-4}\omega_o$ and $r \approx 5a_o$, a lower limit for $r$, and realizing that $\tau\omega_o$ is nearly the third power of the fine structure constant

$$\tau\omega_o = \frac{2q^2/4\pi\varepsilon_0}{3\mu c^3} \frac{\mu(q^2/4\pi\varepsilon_0)^2}{2\hbar^3} = \frac{1}{3}\left(\frac{q^2/4\pi\varepsilon_0}{\hbar c}\right)^3 = \frac{1}{3}\alpha_{sf}^{\,3} \qquad \text{(A-22)}$$

we calculate that $\Gamma_2^{SC}$ is about $10^{-39}\hbar\omega_o$, an extremely small quantity. The equivalent tangential braking force is totally negligible in front of the van der Waals attractive force, since their ratio is $\frac{32}{35\pi}(\tau\omega_o)^2\left(\frac{\Omega}{\omega_o}\right)^7$ which is about $10^{-42}$ if $\frac{\Omega}{\omega_o} \approx 10^{-4}$. Before tackling the self-consistent torque $\Gamma_1^{SC}$ which necessitates harsher calculations, we are going to derive the simple perturbative treatment of (A-6-x) and (A-6-y) and show that the associated torque is $\Gamma_2^{SC}$.

## 4. Perturbative treatment of the equation of motion

A perturbative, first order resolution of (A-6-x) and (A-6-y) yields for the (perturbative) dipolar components:

$$P_x^{\text{pert}}[\omega] = 2\alpha[\omega]\left(E_x[\omega] + \frac{1}{2}\frac{\alpha[\omega]E_x[\omega] + \frac{3}{2}\alpha[\omega+2\Omega](E_x[\omega+2\Omega] - iE_y[\omega+2\Omega]) + \frac{3}{2}\alpha[\omega-2\Omega](E_x[\omega-2\Omega] + iE_y[\omega-2\Omega])}{4\pi\varepsilon_o r^3}\right) \qquad \text{(A-23-x)}$$

$$P_y^{\text{pert}}[\omega] = 2\alpha[\omega]\left(E_y[\omega] + \frac{1}{2}\frac{\alpha[\omega]E_y[\omega] + \frac{3}{2i}\alpha[\omega+2\Omega](E_x[\omega+2\Omega] - iE_y[\omega+2\Omega]) - \frac{3}{2i}\alpha[\omega-2\Omega](E_x[\omega-2\Omega] + iE_y[\omega-2\Omega])}{4\pi\varepsilon_o r^3}\right) \qquad \text{(A-23-y)}$$

The elementary torque (A-15) becomes

$$\gamma^{\text{pert}}[\omega,-\omega,\Omega] = \left\langle P_x^{\text{pert}}[\omega] \cdot \frac{P_y^{\text{pert}}[\omega']}{2\alpha[\omega']} - P_y^{\text{pert}}[\omega] \cdot \frac{P_x^{\text{pert}}[\omega']}{2\alpha[\omega']} \right\rangle \qquad \text{(A-24)}$$

Similarly to the self-consistent case, only $\omega' = -\omega$ yields non non-zero contributions. Using equation (1) and the relation $\frac{\text{Im}[\alpha[\omega]]}{|\alpha[\omega]|^2} = \frac{2}{3}\frac{\omega^3}{4\pi\varepsilon_o c^3}$, the integration of (A-24) yields for the perturbative torque

$$\vec{\Gamma}^{\text{pert}} = i\frac{3}{2}\frac{\hbar}{\pi c^3}\int_{-\infty}^{+\infty}\frac{\alpha[\omega]}{4\pi\varepsilon_o}\left(\left|\frac{\alpha[\omega+2\Omega]}{4\pi\varepsilon_o r^3}\right|^2|\omega+2\Omega|^3 - \left|\frac{\alpha[\omega-2\Omega]}{4\pi\varepsilon_o r^3}\right|^2|\omega-2\Omega|^3\right)d\omega\,\vec{K}$$

which transforms into

$$\vec{\Gamma}^{\text{pert}} = -2\frac{\hbar}{\pi c^6}\int_0^{+\infty}\frac{|\alpha[\omega]|^2}{(4\pi\varepsilon_o)^2}\omega^3\left(\left|\frac{\alpha[\omega+2\Omega]}{4\pi\varepsilon_o r^3}\right|^2|\omega+2\Omega|^3 - \left|\frac{\alpha[\omega-2\Omega]}{4\pi\varepsilon_o r^3}\right|^2|\omega-2\Omega|^3\right)d\omega\,\vec{K} \qquad \text{(A-25)}$$

After the same change of variable as applied to (A-19), we finally get that $\vec{\Gamma}^{\text{pert}} = \Gamma_2^{SC}\vec{K}$ at the lowest order in $\frac{a_o}{r^3}$. The perturbative treatment of dipoles interaction yields a very small and accelerating torque only. This torque has a negligible impact on the atomic motion, but is physically unsatisfactory. The self-consistent treatment is thus a necessity.

## 5. Detailed calculation of the self-consistent torque [4]

Equation (A-18-1) is further worked out:

$$\Gamma_1^{SC} = -2\frac{2}{3}\frac{2\hbar}{3\pi c^3}\int_0^{+\infty}\frac{|\alpha_{xy}[\omega]|^2}{(4\pi\varepsilon_o)^2}\omega^6\left(\frac{1}{\left|1-\frac{9}{4}\frac{\alpha_{xy}[\omega+2\Omega]\alpha_{xy}[\omega]}{(4\pi\varepsilon_o r^3)^2}\right|^2} - \frac{1}{\left|1-\frac{9}{4}\frac{\alpha_{xy}[\omega-2\Omega]\alpha_{xy}[\omega]}{(4\pi\varepsilon_o r^3)^2}\right|^2}\right)d\omega \quad \text{(A-26)}$$

At $\omega = \omega_{xy}$, $\alpha_{xy}[\omega]$ peaks at $\frac{\alpha_{xy}[0]}{\tau\omega_{xy}} \cong \frac{\alpha[0]}{\tau\omega_0}\left(\frac{\omega_o}{\omega_{xy}}\right)^3$, which is $\frac{1}{\tau\omega_0} = \frac{1}{\beta}$ bigger than the value at $\omega = 0$. Thus, in order to use the smallness of β (about $10^{-7}$), we change variables for dimensionless ones. We take $\beta' = \tau\omega_{xy}$ as small parameter. Then we consider $Z = \frac{2(\omega-\omega_{xy})}{\beta'\omega_{xy}}$, $\alpha_o' = \frac{q^2}{4\pi\varepsilon_0\mu\omega_{xy}^2}$, $\xi = \frac{\alpha_o'}{r^3}$ and $\zeta = \frac{\Omega}{\omega_{xy}}$. It is necessary to consider Z and not simply $\frac{\omega-\omega_{xy}}{\omega_{xy}}$ as a dimensionless variable, in order to avoid divergences in the Taylor series in β' (to get rid of β' at the denominator, it is also possible to consider the secondary variables $X = \frac{\xi}{\beta'}$ and $\delta = \frac{\zeta}{\beta'}$). With the new integration variable Z, we write

$$\Gamma_1^{SC} = -2\frac{2}{3}\frac{2\hbar}{3\pi c^6}\left(\frac{\alpha_o'}{\beta'}\right)^2\omega_{xy}^7\int_{-\frac{2}{\beta'}}^{+\infty}\frac{\left(1+\frac{\beta'Z}{2}\right)^6}{\left(Z^2\left(1+\frac{\beta'Z}{4}\right)^2+\left(1+\frac{\beta'Z}{2}\right)^6\right)}\left(f[\beta',Z,\xi,\zeta]-f[\beta',Z,\xi,-\zeta]\right)\frac{\beta'}{2}dZ \quad \text{(A-27)}$$

with

$$f[\beta',Z,\xi,\zeta] = \frac{1}{\left|1-\frac{9}{4}\frac{\alpha_{xy}[\omega+2\Omega]\alpha_{xy}[\omega]}{(4\pi\varepsilon_o r^3)^2}\right|^2}$$

$$= \frac{1}{\left|1-\frac{9}{4}\left(\frac{\xi}{\beta'}\right)^2\left(\frac{1}{\left(-\left(Z+4\frac{\zeta}{\beta'}\right)\left(1+\frac{\beta'Z}{4}+\zeta\right)-i\left(1+\frac{\beta'Z}{2}+2\zeta\right)^3\right)\left(-Z\left(1+\frac{\beta'Z}{4}\right)-i\left(1+\frac{\beta'Z}{2}\right)^3\right)}\right)\right|^2}$$

From the structure of the integrand of (A-27), we see that the first terms in Taylor series in ζ and ξ should scale as $\zeta^1$ and $\xi^2$. This behavior was checked by direct numerical integration of (A-27), up to the highest and still relevant values of the parameters (ζ ≤ $10^{-3}$, ξ ≤ $10^{-2}$). Also, it is clear that the integrand in (A-27) takes non negligible values in the vicinity of Z = 0 only. Thus the lower integration bound can be extended to -∞, with only a small correction discussed later. The integral (A-27) becomes:

---

[4] Algebraic and numerical computations were performed with extensive use of Mathematica software.

$$\Gamma_1^{SC} = -\frac{\hbar\omega_{xy}}{\pi}\tau\omega_{xy}\int_{-\infty}^{+\infty}\frac{\left(1+\frac{\beta'Z}{2}\right)^6}{\left(Z^2\left(1+\frac{\beta'Z}{4}\right)^2+\left(1+\frac{\beta'Z}{2}\right)^6\right)}\left(f[\beta',Z,\xi,\zeta]-f[\beta',Z,\xi,-\zeta]\right)dZ \tag{A-28}$$

After taking the first term of the Taylor series in $\zeta$ and $\xi$, we develop the integrand in $\beta'$ up to the zeroth $[\beta'^0]$ order. As expected, we obtain non zero terms in $\frac{\zeta\xi^2}{\beta'^3}$, $\frac{\zeta\xi^2}{\beta'^2}$ and $\frac{\zeta\xi^2}{\beta'}$. However, their integration on Z from $-\infty$ to $+\infty$ yields an exactly null contribution. We are left with the term in $[\beta'^0]$.

$$\Gamma_1^{SC} \cong -\frac{\hbar\omega_{xy}}{\pi}\tau\omega_{xy}\zeta\xi^2\int_{-\infty}^{+\infty}\left(-\frac{9\left(\left(-672Z^2+14514Z^4-38855Z^6+22895Z^8-3029Z^{10}+35Z^{12}\right)\right)}{16(1+Z^2)^7}\right)dZ \tag{A-29}$$

The algebraic integral equals $\frac{9}{8}\pi$, which gives the result :

$$\Gamma_1^{SC} \cong -\frac{9}{8}\hbar\omega_{xy}\tau\omega_{xy}\zeta\xi^2 = -\frac{9}{8}\hbar\omega_{xy}\tau\omega_{xy}\frac{\Omega}{\omega_{xy}}\left(\frac{\omega_o^2\alpha_o}{\omega_{xy}^2 r^3}\right)^2 = -\frac{9}{8}\hbar\Omega\frac{\omega_o^3}{\omega_{xy}^3}\tau\omega_o\left(\frac{\alpha_o}{r^3}\right)^2 \cong -\frac{9}{8}\hbar\Omega\tau\omega_o\left(\frac{\alpha_o}{r^3}\right)^2\left(1+\frac{3}{4}\frac{\alpha_o}{r^3}\right) \tag{A-30}$$

The lowest order term of (A-30) yields (9), the main result of this work. Numerical integration of (A-28) shows that (A-30) is valid on an extended range of parameters $(\xi,\zeta)$. Also, numerical calculations yield a negligible contribution from the over-counted integration range (from $-\infty$ to $-2/\beta'$).